\documentstyle[12pt]{article}

\begin{document}

\def\kms{km s$^{-1}$}
\def\vlsr{$V_{\rm lsr}$}
\def\Msun{M_\odot}

\title{Galactic Center Shells and a Recurrent Starburst Model}

\author{Yoshiaki S{\sc OFUE}\\
\\
Institute of Astronomy, The University of Tokyo, Mitaka,  \\
Tokyo 181-0015
}

\date{}
\maketitle

\begin{abstract}

By applying filtering techniques to remove straight filaments
in the 20-cm VLA radio image of the Galactic Center Arc region, we have shown
that numerous concentric radio shells of radii 5 to 20 pc are surrounding
the Pistol and Sickle region, which we call Galactic Center Shells (GCS).
Each shell has thermal energy of the order of $10^{49--50}$ erg.
Several CO-line shells are associated, whose kinetic energies
are of the order of $10^{49-50}$ erg.
Summing up the energies of recognized GCSs, the total energy amounts to
$\sim 10^{51}$ erg.
The GCSs show an excellent correlation with the FIR shells
observed at 16--26 $\mu$m with the MSX.
We propose a model in which GCSs were produced by
recurrent and/or intermittent starbursts in the Pistol area during the
last million yr.
The most recent burst occurred some $10^5$ years ago, producing an inner
round-shaped shell (GCS I);  earlier ones a million years ago
produced outer shells (GCS II and III), which are more deformed
by interactions with the surrounding ISM and Sgr A halo.
We argue that recurrent starbursts had also occurred in the past,
which produced larger scale hyper shell structures as well.
A burst some million years ago produced the Galactic Center Lobe,
and a much stronger one 15 million years ago produced the North Polar Spur.
\end{abstract}

\section{Introduction}

The Galactic Center radio-continuum Arc comprises a bunch of  highly aligned
magnetic fields vertical to the galactic plane.
Various thermal features, such as the ``Sickle" and ``Bridge", are superposed
(Yusef-Zadeh, Morris 1987a,b).
Besides the prominent features, several fainter loops with larger
diameters are known to be apparently superposed on the Arc
(Yusef-Zadeh, Morris 1987b).
However, since the radio emission from the Arc dominates,
the faint radio features have not been investigated in details.
A high-resolution CO-line survey has revealed various-scale 
molecular shells in the direction of the Radio Arc (Oka et al. 1998).

In the present work, we tried to enhance the concentric shell structures by
applying a filtering technique, which removed straight filaments
in the Radio Arc. We showed that the enhanced features comprise
numerous thermal radio shells, which we call the Galactic-Center Shells
(GCS), in positional coincidence with 16--26 $\mu$m observed
with the MSX (Shipman et al. 1997).
We also compared the shells with CO-line morphology around the Arc regions
using a high-resolution CO survey of the Galactic Center (Oka et al. 1998),
while a direct interaction of the CO clouds with the Radio Arc is still
controversial.

Based on the obtained images, we propose a new idea
that the GCSs are coherent expanding fronts from a starburst in the GC
near the Pistol and Quintuplet stars.
Starbursts in the Galactic Center (GC) are among the major interesting topics
in the history of the activity of the Milky Way.
We discuss the GCS as being a manifestation of the most recent starburst
in the GC, where past recurrent bursts yielded various scale shells, such as
the North Polar Spur (NPS: Sofue 1977, 1994, 2000) and
the Galactic-Center Lobe (GCL: Sofue, Handa 1984).
We further propose a recurrent starburst model for the GCS, GCL and
NPS.

\section{Filtering Technique and Data}

We used the VLA 20 and 6-cm radio data from Yusef-Zadeh and Morris (1987a,b)
with angular resolutions of 16$''$ and 2$''$, respectively,
which are available as an archive in FITS format and are distributed
on a CD-ROM (Condon, Wells 1992).

We first applied the radial-relieving method (Sofue 1993), while taking the center
at the Sickle (G$0.18-0.04$) position.
The radial-relieving method comprises the following procedure.
A radio image is slightly enlarged,  e.g. by a factor of 1.05,
concentric to a position which is supposed to be the center of
a shell to be enhanced.
The original map is then subtracted from the enlarged
map with the center positions coinciding.
The difference map gives a relieved residual, enhancing
loop-like features concentric to the center position, while other
extended features are suppressed.
 We also used the pressing method (Sofue, Reich 1979) in order
to subtract the straight filaments in the Arc, which is
the same procedure to as that used remove scanning effects from a radio map,
except that the scan-direction is assumed to be parallel to the
filaments. We also used  a background-filtering technique
(Sofue and Reich 1979) in order to subtract the extended background
emission, such as that due to the Sgr A halo.

Among the various features abstracted from the data, we have taken
those features to be real, only if the amplitudes were significantly
greater (e.g. more than 3 times) than the r.m.s. noise of the original image,
and the features could be recognized on the original maps as well.
Here, the r.m.s. noise was estimated for emissions from the most quiet
region in the original map, including the residual interferometer
patterns. The radial relieving technique has the potential to enhance
artificial concentric patterns.
In order to confirm that the enhanced features are real, we 
cross-checked the result from three different methods:
(a) radial relieving to enhance the shell (loop)
features, (b) pressing method to remove straight filaments,
and (c) background filtering (unsharp masking).
The features discussed here are all visible in the three results.
They were also confirmed to be visible in the original maps,
if their intensities are properly displayed individually.

\section{Multiple Radio Shells}

\subsection{Open Lotus}

The results are shown in figures 1 and 2 in galactic coordinates.
Figure 1a shows a radial-relieved  20-cm image (original data from
Yusef-Zadeh and Morris 1987b); Figure 1b is a pressed image, where
the vertical filaments have been removed by the pressing method.
Figure 2a shows the same as figure 1 in a contour form, and figure 2b
is the same but in grey scale and smoothed to an angular resolution 
of 20$''$, where we illustrate the positions of the identified 
radio shells. We name them GCS (Galactic Center Shells) I, II and III.
We also mark the Sickle (Yusef-Zadeh and Morris 1987a,b) by S.

	\centerline{--- Fig. 1a, b, 2a, b ---}

These loops are clearly recognized on the original 20-cm image as well as on the
6-cm images.
The fact that the same loop features are recognized in the original 6 and
20-cm maps confirms that the loops are not an artifact of the reduction
procedure.
Note, however, that because their faintest parts are contaminated by various
interferometer patterns,  we discuss only the global features here.

The most pronounced loop, GCS I,  is centered on G$0.170-0.125$
at $(l,b)=(0^\circ.170, -0^\circ.125)$ with a radius
of $4'.24$, which comprises an almost perfectly round loop.
At an assumed distance of the Galactic Center of 8 kpc, this radius
corresponds to 9.9 pc.
The Sickle (S) at G0.192-0.62 with a radius of $0'.95$
is apparently in touch with GCS I at the north-western inner edge, and
the 'Handle' appears to compose a part of GCS I.
The brightest loops, Shells II and III,
are coincident with the thermal filaments in the Radio Bridge,
which are also concentric to the other loops.
In addition to these prominent loop features, there are several
segments of loops, or arcs, which are concentric to each other
with their centers near to the Sickle and Pistol.
They do not necessarily make  perfect loops, but are sometimes
oval and partial.

Because of their round shapes, these loops are most likely
tangential views of multiple shells.
The shells  appear to compose a coherent structure, suggesting
multiple expanding spherical fronts concentric to the Pistol/Sickle region.
The galactic-western (negative-longitude) sides of the shells are 
apparently contacting the halo of Sgr A, and are much brighter than the 
opposite side.
The galactic-eastern sides are less bright, and seem to be expanding 
more freely with less deformation.
As a whole, the shells look like open petals of a lotus bloom
with its neck at the Sgr A halo (figures 2a, b).
It is interesting to note that the shells appear to have no
clear indication of any interaction with the straight nonthermal
filaments in the Radio Arc.
In fact, GCS I, the almost perfectly round shell, is not deformed by
the Radio Arc.

\subsection{Properties of Radio Emission}

Most of the GCS are visible in the 43-GHz map (Sofue et al. 1986).
The spectral index of the NE part of GCS II, inferred from
the 43 and 1.4 GHz intensities, is about $\sim -0.05$,
indicating a thermal origin due to ionized interstellar gas.
The typical brightness temperature on GCS II is $\sim 50$ K at 20 cm,
which yields an emission measure of $\sim 1.2 \times 10^5$ pc cm$^{-6}$,
if the electron temperature is taken to be $\sim 10^4$ K.
Assuming that the thickness of the shell is 0.1 times the
radius ($0.1 \times 9.9$ pc $\sim 1$ pc), the line-of-sight depth
will be about $\sim 4$ pc.

This yields an electron density of $\sim 1.7 \times 10^2$ cm$^{-3}$,
and a total mass of the ionized hydrogen $\sim 5\times 10^3 \Msun$.
The thermal energy would, then, be $ \sim 1 \times 10^{49}$ erg.
However, the southern part appears to have a steeper spectrum as
inferred from 20 and 6 cm maps, although the present data are not
sufficiently accurate to derive spectral indices in such faint features.
Thus, we cannot exclude the possibility that the shell comprises nonthermal
emission, such as a supernova remnant.

\subsection{Far-Infrared Shells}

GCS I coincides in positions with the most prominent far-infrared loop
observed by the MSX experiment at 16--26 $\mu$m (Shipman et al. 1997).
The MSX image also shows many other shells, coincident with the other
GCSs.
In figures 3 we compare the radio shells with the MSX image.
The association of the FIR emission may indicate that the shells
are predominantly thermal, consistent with the radio spectrum, and
contain warm dust and probably molecular gas.
It is interesting to note that the Arc is not visible in the FIR at all.
Thus, the MSX FIR image appears to be very similar to the radio image
after subtracting of the Arc filaments.

	\centerline{--- Fig. 3 ---}

\subsection{Molecular Shells}

The association of molecular gas with the radio Arc and Bridge has been
discussed in details by Serabyn and Guesten (1987).
Numerous CO-line shells and arcs are found toward
the presently identified GCSs, whereas no clear CO feature
is associated with the Arc.
Figure 4a shows an unsharp-masked total CO intensity map, and figure 4b
is a superposition on the radio continuum image from figure 1a.
Figure 5 shows channel maps of the CO emission.
Here, the CO map has been produced using the data cube of the Galactic
Center CO survey by Oka et al. (1998).
In figure 4a, we indicate GCS I, II, III, and the Sickle (S) by full lines. 

\centerline{--- Fig. 4, 5 ---}

Oka et al. (1998) noticed an expanding shell at 50 \kms\ with a radius
and thickness of $\sim 9$ and  $\sim 4$ pc, respectively, 
centered on $(l, b)= (15',-4')$. 
We illustrate this shell by the  dotted circle in figures 4a and 5.
The CO intensity is roughly $\sim 10^2$ K \kms,
yielding a molecular mass of $2 \times 10^4 \Msun$
for a conversion factor of
$1.0 \times 10^{20}$ H$_2$ [K \kms]$^{-1}$ (Arimoto et al. 1996).
The shell is expanding at 25 \kms,  which yields a kinetic energy
of $\sim 1.4\times10^{50}$ erg, and an age of $\sim 3.5\times 10^5$ yr.

Besides this expanding shell, there are several arc-features in the channel
maps, some of which appear to be spatially coincident with the GCS positions.
GCS I appears to be associated with a CO arc  at 10 to 30 \kms.
GCS II is associated with a dense CO arc at 60 to 70 \kms.
The galactic-northern part of GCS III is associated with an extended CO arc
at $-10$ to $-20$ \kms.
However, no conclusive physical association can be derived from these
coincidences at the present resolution, and we cannot exclude the possibility
of a chance coincidence.

\subsection{X-ray Sources in GCS I}

The ASCA Galactic Center survey has revealed several bright X-ray sources
in the 6.4 keV iron-line (Maeda 1998).
The brightest source coincides with Sgr B.
The second-strongest source coincides with the Radio Bridge,
with which GCS II and III are associated.
Most interestingly, the third-strongest source, associated with the
Arc region,  coincides with the center of GCS I, which is toward the hole
in the FIR 16--26 $\mu$m emission.
An extended continuum X-ray source at 0.7--10 keV is also found near to the 
center of GCS I.

\section{Discussion}

\subsection{GCS Energetics}

The concentric distribution of the GCSs suggests that they have
a common origin, most likely due to mass loss or explosions associated
with active star formation near to the Pistol and Sickle.
In fact, intense star-forming activity has been observed as the Pistol
star and Quintuplet stars (Figer et al. 1998, 1999; Lang et al. 1999).
Figer et al. (1999) estimate the mass-loss rate and duration to be
$10^{-4} \Msun~{\rm yr^{-1}}$ and 6000 yr with a wind velocity of
2000 \kms, which yields a total kinetic energy of $\sim 2.4\times10^{49}$ erg.
This is comparable to the thermal energy of a single GCS and the kinetic
energy of an associated CO shell.

If such giant stars, or an ensemble of massive stars, like the
Quintuplet, were born recurrently in the past, they would have produced
multiple ionization shells, which expanded to sizes as large as several
to tens of parsecs.
Interactions with and/or the accumulation of background interstellar gas and
molecular clouds would decelerate the expansion velocity to several tens of \kms,
as observed in the GCS.
Hence, the GCSs would be shells in the evolved phases of similar nebulae to the
Pistol.
If the total energy $E$ is conserved, the expansion velocity $v$ and
radius $r$ are related as $E\sim {2\pi \over 3} \rho r^3 v^2$,
where $\rho$ is the background gas density.
Taking $E\sim 2.4\times10^{49}$ erg,  and $v \sim 25$ \kms, the averaged ISM
density is required to be $\rho\sim 20 ~{\rm H}_2~ {\rm cm}^{-3}$.
The age is of the order of $t\sim r/v \sim 4\times 10^5$ yr.

The total energy of the Pistol, Sickle and all the GCSs identified here is
of the order of $10^{51}$ erg.
The age of the outermost shell (GCS III) is estimated to be
$\sim 10^6$ years from their radii of $9' \sim 20$ pc, 
and the assumed expansion velocity is 20 \kms.
We may, thus, suppose that recurrent mini-bursts have occurred
near the Pistol/Sickle region in the past million years.
The burst centers were within several parsecs of the center of GCS II.

\subsection{GCS Morphology}

GCS I is almost perfectly round, suggesting that they have not
been disturbed very much by the surrounding ISM.
However, the outer shells, GCS II and III, are significantly deformed,
particularly on their SW side, where they become concave with respect to the
shell centers.
This fact suggests an interaction with the Sgr A halo.
On the contrary, the galactic-eastern sides of GCS II and III, are more 
open, where the shells appear to be torn off into several segments.
One prominent segment is indicated by an arc in figure 2b.

If we assume that GCS II, III and the other segments (as in figure 2) 
are a coherent structure, they may be parts of a large shell, 
which is elongated in the direction
perpendicular to the galactic plane, with the galactic-western
sides being interrupted
by the Sgr A halo and the eastern sides expanding more freely.
If the GCSs are expanding shells in the GC gas disk, the vertical 
elongation is a natural consequence due to the background gas-density 
distribution.
The lopsidedness observed in most of the GCSs will be due to
interactions with an inhomogeneous ISM, such as the Sgr A halo and neighboring
molecular clouds.

\subsection{GCS Origin: Recurrent Starbursts Model}

We have proposed a bipolar hyper shell (BHS) model due to an intense 
energy release, such as starbursts at the GC in order to explain
the large-scale radio shell structures associated with the GC
(Sofue 1977, 1984, 1994, 2000).
This model can be applied to smaller scale shells, such as the GCSs.
BHS simulations have shown that an expanding shell is round in the early 
phase, when the shell radius is smaller than the disk thickness.
GCS I  may be in such an early phase.
Then, the shell becomes elongated in the direction perpendicular to the disk
when the radius becomes comparable to the disk thickness.
GCS II and III would be in such a phase.
Hence, GCSs  would manifest sequential phases of evolution
of an expanding shell originating around the Pistol.

According to BHS simulations, the shell expands further into the halo,
while becoming more elongated and opened to form a bipolar $\Omega$ shape
or an hour-glass shape.
In fact, radio observations have shown an $\Omega$-shaped
feature above the Galactic Center, called the Galactic
Center Lobe (GCL), which requires energy of the order of $10^{54}$
erg and an age of $\sim 10^6$ yr (Sofue 1985).
If the energy scale is greater, the $\Omega$-shaped shell expands
into the halo, forming a larger scale BHS.
In fact, in our Galaxy an extremely large shell has indeed been observed in
radio and X-rays as the North Polar Spur (NPS), for which the BHS
model has been successfully applied; 
the simulation requires an input energy of $\sim 10^{55}$ erg and an age of
$\sim1.5\times10^7$ yr (Sofue 1977, 1984, 1994, 2000).
Similar BHS or hyper-wind phenomena have also been observed in 
starburst galaxies (Heckman et al. 1990), for which a number of numerical 
simulations have been successfully applied (Tomisaka, Ikeuchi 1988; 
Suchkov et al. 1994; Strickland, Stevens 2000; Sofue, Vogler 2002).

We, thus, propose a recurrent-starburst model in our Galactic Center.
The most recent mini-burst, which occurred some $10^5$ years ago,
produced the inner GCSs,
while the outer GCSs were created by previous ones $\sim 10^{5-6}$ years ago.
Each energy scale of these mini-bursts was of the order of $10^{51}$ erg.
A stronger burst a million years ago produced the Galactic Center Lobe,
for which an energy of $\sim 10^{54}$ erg is required.
A much bigger burst with $\sim 10^{55}$ occurred 15 million years ago, and 
produced the North Polar Spur.
Their energy scales are diverse from $10^{51}$ to $10^{55}$ erg.
The large, long-living shells, like the GCL and NPS, would not
necessarily be due to a single event, but due to the accumulation of
succeeding mini-bursts of GCS scale and bursts of GCL scale, 
respectively. 

\vskip 5mm
The radio continuum data were taken from the
VLA public images, courtesy Drs. F. Yuzef-Zadeh and M. Morris,
distributed by J. J. Condon and D. Wells by a CD-ROM.
The author thanks Dr. T. Oka for the CO-line survey in FITS format.

\vskip 10mm

\noindent{\bf References}
\def\r{\hangindent=2pc  \noindent}

\r Arimoto, N., Sofue, Y., \& Tsujimoto, T. 1996, PASJ, 48, 275

\r Condon, J. J., \& Wells, D. 1992, Images from the Radio Universe
(CD-ROM), NRAO

\r Figer, D. F.,  Najarro, F., Morris, M., McLean, I. S.,
Geballe, T. R., Ghez, A. M., \& Langer, N. 1998, ApJ, 506, 384

\r Figer, D. F., Morris, M., Geballe, T. R.,  Rich, R. M., Serabyn, E.,
McLean, I. S.,  Puetter, R. C., \& Yahil, A. 1999, ApJ, 525, 759

\r Heckman, T. M., Armus, L., \&  Miley, G.  K., 1990, ApJS, 74, 833

\r Lang, C. C., Figer, D. F., Goss, W. M., \&  Morris, M. 1999, AJ, 118, 2327

\r Maeda, Y. 1998, PhD Thesis, Kyoto University
(ISAS Research Note No.  653)

\r Oka, T., Hasegawa, T., Sato, F., Tsuboi, M., \& Miyazaki, A.
1998, ApJS, 118, 455

\r Serabyn, E, \& Guesten, R. 1987, A\&A, 184, 133

\r Shipman, R. F.,  Egan, M. P., \& Price, S. D.  1997,
Galactic Center News, ed. A. Cotera and H. Falcke, Vol 5, 3

\r Sofue, Y. 1977, A\&A, 60, 327

\r Sofue, Y. 1984,  PASJ,  36,  539

\r Sofue, Y. 1985,  PASJ,  37,  697

\r Sofue, Y. 1994, ApJ, 431, L91

\r Sofue, Y. 2000, ApJ, 540, 224

\r Sofue, Y. 1993, PASP, 105, 308

\r Sofue, Y., \& Handa, T. 1984, Nature, 310, 568

\r Sofue, Y., Inoue, M.,  Handa, T., Tsuboi, M.,  Hirabayashi, H.,
Morimoto, M., \&  Akabane, K. 1986, PASJ, 38, 475

\r Sofue, Y., \& Reich, W. 1979, A\&AS, 38, 251

\r Sofue, Y., \& Vogler, A. 2001, A\&A, 370, 53

\r Strickland, D. K., \& Stevens, I. R. 2000, MNRAS, 314, 511

\r Suchkov A.A., Belsara, D. S., Heckman T.M., \& Leitherer C., 1994, 
ApJ, 430, 511

\r Tomisaka, K., \& Ikeuchi, S. 1988, ApJ, 330, 695

\r Yusef-Zadeh, F., \& Morris, M. 1987a, ApJ, 322, 721

\r Yusef-Zadeh, F., \& Morris, M. 1987b, AJ, 94, 1178

\newpage

\parindent=0pt
\parskip 6mm

Figure Captions

Fig. 1a. Radial-relieved image of the Galactic Center Arc region
observed with the VLA at 20 cm (1.446 GHz: original data from
Yusef-Zadeh,  Morris 1987a). Numerous loop features are prominent.

Fig. 1b. Pressed VLA 20-cm map, where straight Arc filaments have been
removed by the pressing method. Extended emissions were also
subtracted by a background-filtering technique.

Fig. 2a. Same as figure 1a in a contour form.
Contours are drawn at 1, 2, 4, 8, 16, 32, 64 mJy/(16$''$ Beam).

Fig. 2b. Same as figure 2a, but smoothed to a resolution of $20''$.
The Galactic-Center Shells I to III and the Sickel (S)  are indicated
by the full lines.
Contours are drawn at 1, 2, 4, 8, 16, 32, 64 mJy/(16$''$ Beam).

Fig. 3. Comparison of the radio GCS (left) with the 
MSX 16--26 $\mu$m image (right: Shipmann et al. 1997).

Fig. 4. Unsharp-masked CO integrated-intensity map produced by using the
CO survey by Oka et al. (1998). Contour levels are at
6.33$\times$(1, 2,..., 10) K \kms.
The GCS I, II, III and the Sickle (S) are indicated by full lines, and the
expanding  shell of Oka et al. is indicated by a dotted circle
(ES+50km).

Fig. 5. CO intensity channel maps from Oka et al. (1998)
at $V_{\rm lsr} =-60$ to 130 \kms\ at 10 \kms\ interval.
The contour levels are at 1.11$\times$(1, 2,..., 10) K \kms.
The GCSs are indicated by full lines, and the
expanding CO shell by the dotted circle.

\end{document}